\documentclass[showpacs,twocolumn]{revtex4}
\usepackage{epsfig}
\usepackage{multirow}
\begin{document}

\title{Determination of factors governing fibrillogenesis of polypeptide chains using lattice models}

\author{Mai Suan Li$^1$, Nguyen Truong Co${^2}$, Govardhan Reddy$^{3}$, C-K. Hu$^{5,6}$, 
and D. Thirumalai$^{3,4}$}

\affiliation{$^1$Institute of Physics, Polish Academy of Sciences,
Al. Lotnikow 32/46, 02-668 Warsaw, Poland\\
$^2$Saigon Institute for Computational Science and Technology,
6 Quarter, Linh Trung Ward, Thu Duc District, Ho Chi Minh
City, Vietnam\\
$^3$Biophysics Program, Institute for Physical
Science and Technology, University of Maryland, College Park, MD 20742\\
$^4$Department of Chemistry and Biochemistry, University of Maryland,
College Park, MD 20742\\
$^5$Institute of Physics, Academia
Sinica, Nankang, Taipei 11529, Taiwan\\
$^6$Center for Nonlinear and Complex Systems and Department of
Physics, Chung Yuan Christian University, Chungli 32023, Taiwan
}

\begin{abstract}
Using lattice models we explore the factors that determine the tendencies of polypeptide chains to aggregate by exhaustively sampling the sequence and conformational space. The morphologies of the fibril-like structures and the time scales ($\tau_{fib}$) for their formation depend on a subtle balance between hydrophobic and coulomb interactions. The extent of population of \textbf{N$^*$}, which is a fibril-prone structure in the spectrum of monomer conformations, is the major determinant of $\tau_{fib}$.  This observation is used to determine the aggregation-prone consensus sequences by exhaustively exploring the sequence space.  Our results provide a basis for genome wide search of fragments that are aggregation prone.
\end{abstract}


\maketitle

Proteins that are unrelated by sequence or structure aggregate to form amyloid-like fibrils with a characteristic cross $\beta$-structures, which are linked to a number of deposition diseases such as Alzheimer's and prion-disorders~\cite{gen_ref}(a). The observation that almost any protein could form fibrils seemed to imply that fibril rates can be predicted solely based on sequence composition and the propensity to adopt global secondary structure - a conclusion that has limited validity. Despite the common structural characteristics of organized aggregates such as amyloid fibrils~\cite{gen_ref}(b)-(e) the factors that determine the fibril formation tendencies are not understood. 

Experiments on fibril formation times ($\tau_{fib}$) have been rationalized using global factors such as the hydrophobicity of side chains~\cite{factor}(a), net charge~\cite{factor}(b,c), patterns of polar and non-polar residues~\cite{factor}(d), frustration in secondary structure elements~\cite{factor}(e,f), and  aromatic interactions~\cite{factor}(g). However, the inability to sample the sequences and conformational spaces exhaustively~\cite{all_atom} has prevented deciphering plausible general principles that govern protein aggregation using limited computations and experiments. The purpose of this letter is to obtain a quantitative correlation between intrinsic properties of polypeptide sequences and their fibril growth rates using the lattice models, which have given remarkable insights into the general principles of protein folding and aggregation~\cite{lattice}. Using a modification of the model in~\cite{MSLi_JCP08} we explore the sequence-dependent variations of  $\tau_{fib}$  on the nature of conformations explored by the monomer. In particular, we highlight the role of {\bf N}$^*$, which is one of the aggregation prone structures~\cite{DT_cosb_03} in the folding landscape of the monomer in determining $\tau_{fib}$ and the propensity of sequences to form fibrils. 

\begin{figure}
\includegraphics[width=3.0in]{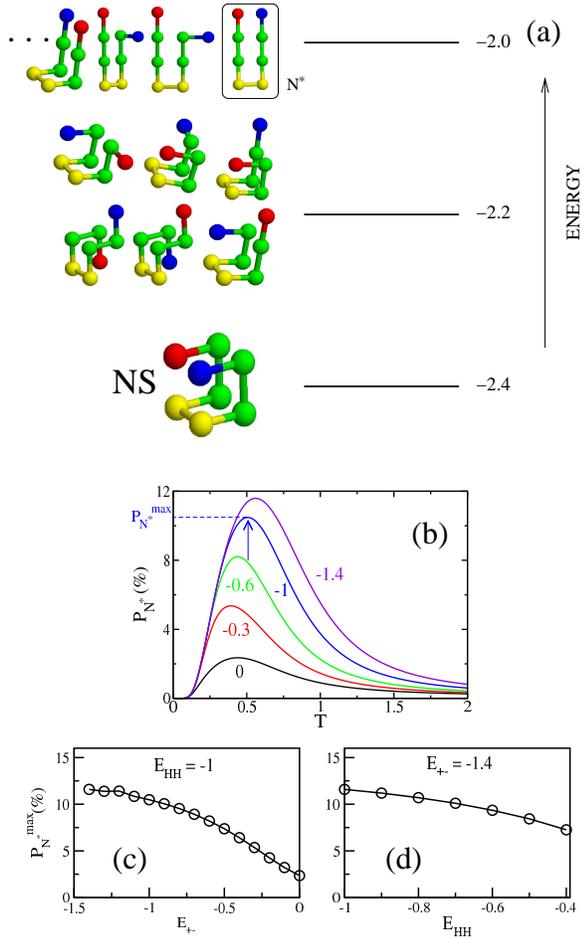}
\caption{(a) Spectrum of energies and associated low energy structures of the monomer sequence +HHPPHH-. H, P, + and - are in green, yellow, blue, and red, respectively. We set $E_{HH}=-1$ and $E_{+-}=0$. There are a total 1831 possible conformations that are  spread among 17 possible energy values. The {\bf N}$^*$ structure enclosed in the box coincides with the peptide state in the fibril (see Fig.~\ref{n6_fib_nonfib_snap_tfn6_q14q06_fig}a). (b) The probability $P_{N^*}$ of populating the aggregation-prone structure {\bf N}$^*$ as a function of $T$ for $E_{+-}=0, -0.3, 0.6, -1$ and -1.4 keeping $E_{HH}=-1$. The arrow indicates $T^*$, where $P_{N^*}=P_{N^*}^{max}$.  Dependence of $P_{N^*}^{max}$ on $E_{+-}$ for $E_{HH}=-1$ (c), and on $E_{HH}$ for $E_{+-}=-1.4$ (d).
}
\label{spectrum_charge_Pn_charge_hp_fig}
\end{figure}

{\bf Lattice model.}
To explore the dependence of fibril formation rates on the intrinsic properties of monomers and the sequence, we use a lattice model~\cite{MSLi_JCP08} in which each chain consists of $M$ connected beads that are confined to the vertices of a cube. The simulations are done using $N$ identical chains with $M=8$. The peptide sequence which is used to illustrate the roles of electrostatic and hydrophobic interaction is +HHPPHH- (Fig.~\ref{spectrum_charge_Pn_charge_hp_fig}), where H, P, + and - are hydrophobic, polar, positively charged and negatively charged beads  respectively~\cite{MSLi_JCP08}.  

The inter- and intra-chain potentials include excluded volume and contact (nearest neighbor) interactions. Excluded volume is imposed by the condition that a lattice site can be occupied by only one bead. The energy of $N$ chains is~\cite{MSLi_JCP08}

\begin{equation}
E  =  \sum_{l=1}^N \, \sum_{i<j}^M \; E_{sl(i)sl(j)} \delta (r_{ij}-a) + \nonumber 
\sum_{m<l}^N \; \sum_{i,j}^M \; E_{sl(i)sm(j)} \delta (r_{ij}-a) ,
\label{energy_eq}
\end{equation}
where $r_{ij}$ is the distance between residues $i$ and $j$, $a$ is a lattice spacing, $sm(i)$ indicates the type of residue $i$ from $m$-th peptide, and $\delta (0)=1$ and zero, otherwise.  The first and second terms in Eq. \ref{energy_eq} represent intrapeptide and interpeptide interactions, respectively.

 The propensity of polar (including charged) residues to be ``solvated''  is mimicked using $E_{P\alpha} =$-0.2 (in the units of hydrogen bond energy $\epsilon_H$), where $\alpha$= P,+,or -. To assess the importance of electrostatic and hydrophobic interactions, we vary either $E_{+-}$ in the interval $-1.4 \le E_{+-} \le 0$ or $E_{HH}$ between -1 and 0. If $E_{+-}$ is varied, we set $E_{HH}=-1$, while if $E_{HH}$ is varied, then $E_{+-}=-1.4$. We used $E_{++} = E_{--} = -E_{+-}/2$ and all other contact interactions have $E_{\alpha\beta}=$ 0.2.

{\bf Monomer spectra depends on $E_{+-}$ and $E_{HH}$.}
The spectrum of energy states of the monomer for a given sequence is determined by exact enumeration of all possible conformations (Fig.~\ref{spectrum_charge_Pn_charge_hp_fig}). For all sets of  contact energies chosen above the native state (NS) of the monomer is compact (lowest energy conformation in Fig.~\ref{spectrum_charge_Pn_charge_hp_fig}a). In anticipation of the role {\bf N}$^*$ (the structure encircled in the box in Fig.~\ref{spectrum_charge_Pn_charge_hp_fig}a) plays in promoting fibril formation we focus on the change in the rank order of the {\bf N}$^*$ energy level as $E_{+-}$ is varied. For $E_{+-}<0$, {\bf N}$^*$ is the first excited state~\cite{MSLi_JCP08}. However, if $E_{+-}=0$,  {\bf N$^*$} is one of the 19-fold degenerate states in the second excited state (Fig.~\ref{spectrum_charge_Pn_charge_hp_fig}a). The energy gap  between the monomeric NS and state {\bf N$^*$}, $\Delta E=E_{N^*}-E_{NS}=0.4$ for the sequences in Fig.~\ref{spectrum_charge_Pn_charge_hp_fig}, is a constant.

The population of the putative fibril-prone conformation in the monomeric state is $P_{N^*}=\exp(-E_{N^*})/Z$, where $Z$ is the partition function is obtained by exact enumeration.  Fig.~\ref{spectrum_charge_Pn_charge_hp_fig}b, shows the temperature dependence of $P_{N^*}$ for various values of ($E_{+-}$) interaction, with $E_{HH}=-1$ and other contact energies constant. Depending on $E_{+-}$, the maximum value of $P_{N^*}$ varies from 2\% $< P_{N^*}^{max} < $ 12\% (Fig. \ref{spectrum_charge_Pn_charge_hp_fig}c).  $P_{N^*}^{max}$ decreases to a lesser extent as the hydrophobic interaction grows (Fig. \ref{spectrum_charge_Pn_charge_hp_fig}d). Here we consider only $E_{HH} \le -0.4$ because the fibril-like structure is not the lowest-energy when $E_{HH} > -0.4$ .

\begin{figure}
\includegraphics[width=3.0in]{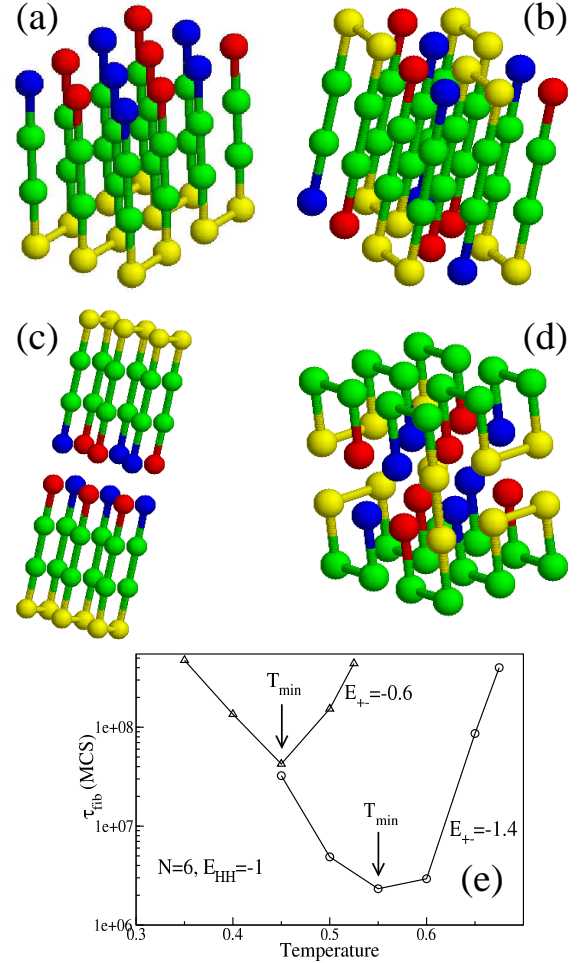}
\caption{(a) The lowest energy fibril structure for $E_{+-}=-1.4$ and 
$E_{HH}=-1$. (b) Same as in (a) but with $E_{+-}=0$. (c) Double layer structure for $E_{HH}=-0.4$ but keeping  $E_{+-}=-1.4$. (d) For $E_{+-}=-1.4$ and $E_{HH}=-0.3$ the fibril structure is entirely altered. (e) Temperature dependence of $\tau_{fib}$ for $E_{+-}=-1.4$ (circles) and $E_{+-}=-0.6$ (triangles). $N=6$ and $E_{HH}=-1$. Arrows show the temperatures at which the fibril formation is fastest.
}
\label{n6_fib_nonfib_snap_tfn6_q14q06_fig}
\end{figure}

{\bf Morphology of lowest-energy structures of multi-chain systems depends on sequences.}
When multiple chains are present in the unit cell, aggregation is readily observed, and in due course they lead to ordered structures. We used the Monte Carlo (MC)~\cite{MSLi_JCP08} annealing protocol, which allows for an exhaustive conformational search, to find the lowest energy conformation. For non-zero values of $E_{+-}$ the chains adopt an antiparallel arrangement in the ordered protofilament, which ensures that the number of salt-bridge and hydrophobic contacts are maximized (Fig. \ref{n6_fib_nonfib_snap_tfn6_q14q06_fig}a and see also Ref.~\cite{MSLi_JCP08}). If $E_{+-}=0$ then the lowest energy fibril structure has a vastly different architecture even though they are assembled from {\bf N}$^*$ (Fig. \ref{n6_fib_nonfib_snap_tfn6_q14q06_fig}b). The structure in Fig.~\ref{n6_fib_nonfib_snap_tfn6_q14q06_fig}b, in which a pair of {\bf N}$^*$ conformations are stacked by flipping one with respect to the other is rendered stable by maximizing the number of +P and -P contacts. We now set $E_{+-}=-1.4$ and vary $E_{HH}$. For $E_{HH}<-0.4$, the fibril conformation adopts the same shape as that shown in Fig. \ref{n6_fib_nonfib_snap_tfn6_q14q06_fig}a, but for $E_{HH}=-0.4$ the energetically more favorable double-layer structure emerges (Fig. \ref{n6_fib_nonfib_snap_tfn6_q14q06_fig}c). If $E_{HH}\ge -0.3$, then the lowest-energy conformation ceases to have the fibril-like shape (Fig. \ref{n6_fib_nonfib_snap_tfn6_q14q06_fig}d). The close packed heterogenous structure is stitched together by a mixture of the NS conformation and one of the second excited conformations. Even for this simple model a  variety of lowest-energy structures of oligomers and protofilaments with different morphologies emerge, depending on a subtle balance between electrostatic and hydrophobic interactions.

{\bf Dependence of $\tau_{fib}$ on $E_{+-}$ and $E_{HH}$.}
 We use MC algorithm to study the kinetics of fibril assembly. Simulations were performed by enclosing $N$ chains in a box with periodic boundary conditions~\cite{MSLi_JCP08}. The monomer concentration was kept $\approx 6$ mM (roughly the length of the cubic size is 35$a$ for $N=10$ monomers) for all systems. The fibril formation time $\tau_{fib}$ is defined as an average of first passage times needed to reach the fibril state with the lowest energy starting from initial random conformations. For a given value of $T$, we generated 50-100 MC trajectories to obtain reliable estimates of $\tau_{fib}$. MC moves include global and local ones. A local move~\cite{move} corresponds to tail rotation, corner flip, and crankshaft rotation. Global moves correspond to either translation of a peptide by $a$ in a randomly chosen direction or rotation by $90^o$ around one of the randomly chosen coordinate axes. The acceptance probabilities of global and local moves are 0.1 and 0.9, respectively~\cite{MSLi_JCP08}. We measure time in units of Monte Carlo steps (MCS). The combination of local and global moves constitutes one MCS.

\begin{figure}
\includegraphics[width=3.0in]{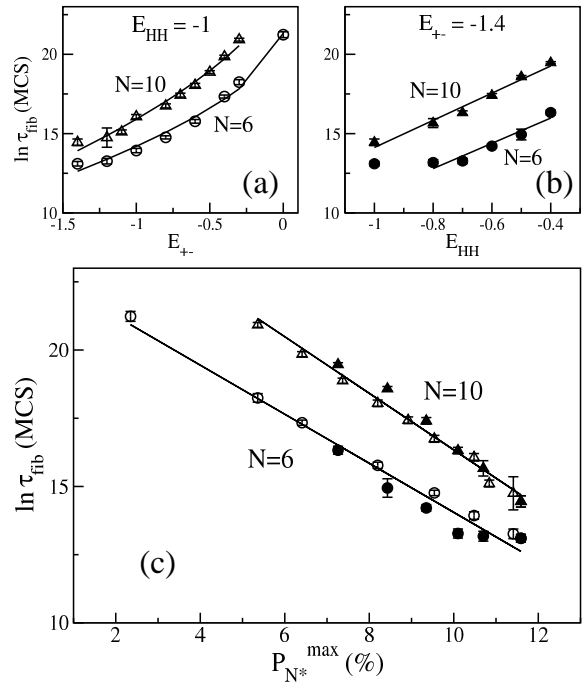}
\caption{(a) Dependence of $\tau_{fib}$ on $E_{+-}$ for $N=6$ (circles) and $N=10$ (triangles) with $E_{HH}=-1$. The solid curves are fits to $y = c_0 + c(-x)^{\alpha}$, where $\alpha \approx 0.59$.
$c_0=21.32$ and $c=-7.12$ and $c_0=25.14$ and $c=-9.23$ for $N=6$ and 10,
respectively. (b)  Dependence of $\tau_{fib}$ on $E_{HH}$ with
$E_{+-}=-1.4$ hold constant for $N=6$ (solid circle) and $N=10$
(solid triangles). Lines are fits $y = 19.17 + 7.97x$
and $y = 22.69 + 8.56x$ for $N=6$ and 10,
respectively. For $N=6$ the first point $E_{HH}=-1$ is excluded from fitting.
(c) Dependence of $\tau_{fib}$ on $P_{N^*}^{max}$ for $N=6$ and 10.
Symbols are the same as in (a) and (b)
$\tau_{fib}$ is measured in MCS and $P_{N^*}^{max}$ in \%. The correlation coefficient for all fits $R \approx 0.98$.}
\label{tf_n6_n10_qh_pol_total_fig}
\end{figure}

The temperature dependence of $\tau_{fib}$ displays a U-shape (Fig.~\ref{n6_fib_nonfib_snap_tfn6_q14q06_fig}e) and the fastest assembly occurs at $T_{min}$, which roughly coincides with the temperature, $T^*$, where $P_{N^*}$ reaches maximum (Fig.~\ref{spectrum_charge_Pn_charge_hp_fig}b). To probe the correlation between $\tau_{fib}$ and $E_{+-}$ and  $E_{HH}$  we performed simulations at $T_{min}$. The dependence of $\tau_{fib}$ on $E_{+-}$ can be fit using $\tau_{fib} \sim \exp[-c(-E_{+-})^{\alpha}]$ where $\alpha \approx 0.6$ and the constant $c \approx 7.12$ and 9.23 for $N=6$ and 10 respectively (Fig.~\ref{tf_n6_n10_qh_pol_total_fig}a). Thus, variation of $E_{+-}$ drastically changes not only the morphology of the ordered protofilament (Fig.~\ref{n6_fib_nonfib_snap_tfn6_q14q06_fig}), but also $\tau_{fib}$. As the strength of the charge interaction between the terminal beads increases, the faster is the fibril formation process. Interestingly, the fibril formation rate at $E_{+-}=0$ is about four orders of magnitude slower than that at $E_{+-}=-1.4$. Our model shows that, in agreement with previous studies~\cite{gen_ref}(f), the propensity to fibril assembly strongly depends on the charge states of the polypeptide sequences.

By fixing $E_{+-}=-1.4$ we calculated the dependence of $\tau_{fib}$ on the hydrophobic interaction (Fig.~\ref{tf_n6_n10_qh_pol_total_fig}b), which may be approximated using $\tau_{fib} \sim \exp(cE_{HH})$. Here constant $c \approx 7.97$ and 8.56 for $N=6$ and 10, respectively.  For $N=10$, a change in hydrophobicity of  $\Delta E_{HH}=0.6$, leads to self-assembly rates that are more than two orders of magnitude. In accord with experiments~\cite{gen_ref, Bowerman_MolBioSys09}, our finding shows that the enhancement of hydrophobic interactions speeds up fibril formation rates.

{\bf Fibril formation rates depend on $P_{N^*}$.} 
The dramatic variations in $\tau_{fib}$ on $E_{+-}$ and $E_{HH}$ prompted us to link the underlying spectrum of monomer conformations to $\tau_{fib}$. 
To establish such a relation we collected all the data in Fig. \ref{tf_n6_n10_qh_pol_total_fig}a and \ref{tf_n6_n10_qh_pol_total_fig}b and plot them as a function of $P_{N^*}^{max}$ (Fig. \ref{tf_n6_n10_qh_pol_total_fig}c).  
We found the surprising relation
\begin{equation}
\tau_{fib} = \tau_{fib}^0\exp(-cP_{N^*}^{max}),
\label{tf_Pn_eq}
\end{equation}
where the prefactor $\tau_{fib}^0 \approx 1.014\times 10^{10}$ MCS and 3.981$\times 10^{11}$ MCS, and $c \approx 0.9$ and 1.0, for $N=6$ and 10, respectively.  There are a few implications of the central result given in Eq.~\ref{tf_Pn_eq}. (i) The sequence-dependent spectrum of the monomer is a harbinger of fibril formation. In proteins there are multiple {\bf N}$^*$ conformations corresponding to distinct free energy basins of attraction~\cite{DT_cosb_03}(b). Aggregation from each of the structures in the various basins of attraction could lead to fibrils with different morphologies (polymorphism) that cannot be captured using lattice models. (ii) Enhancement of $P_{N^*}$ either by mutation or chemical cross linking should increase fibril formation rates. Indeed, a recent experiment~\cite{Sciarretta_Biochemistry05} showed that the aggregation rate of A$\beta_{1-40}$-lactam[D23-K28], in which the residues D23 and K28 are  chemically constrained by a lactam bridge, is nearly a 1000 times greater than in the wild-type. Since the salt bridge constraint increases the population of the {\bf N}$^*$  conformation in the monomeric state~\cite{Reddy_JPCB09}. It follows from Eq.~\ref{tf_Pn_eq}, $\tau_{fib}$ should decrease. (iii) Since  $P_{N^*}(T)$ depends on the spectrum of the precise sequence, it follows that the entire free energy landscape of the monomer~\cite{DT_cosb_03}(b) and not merely the sequence composition as ascertained else where\cite{gen_ref}(f), should be considered in the predictions of the amyloidogenic tendencies of a particular sequence. 

{\bf Scanning the sequence space.} We further exploit the result in Eq.~\ref{tf_Pn_eq}  to determine the amylome~\cite{Eisenberg_pnas10}, the universe of sequences in the lattice model, that can form fibrils. We posit that aggregation prone sequences are those with a unique native state with a maximum in $P_{N^*}(T)$ in the interval $1.0 \le T^*/T_F \le 1.25$. Out of the 65,536 sequences only 217 satisfy these criteria (see Supplementary Information (SI) for details). The sequence space exploration shows that there is a high degree of correlation between the positions of charged and hydrophobic residues leading to a limited number of aggregation prone sequences with +HHPPHH- being an example. In addition, there are substantial variations in $T^*/T_F$ for sequences with identical sequence composition, which reinforces the recent finding~\cite{Eisenberg_pnas10} that context in which charged and hydrophobic residues are found is important in the tendency to form amyloid-like fibrils. 

We have shown that there is a strong correlation between the extent of population of {\bf N}$^*$ structures (in proteins we expect multiple aggregation prone conformations), which depend on the exact sequence and the environment. 
Interestingly, $P_{N^*}^{max}$, which roughly anti-correlates with $\Delta (=[E_{NS}-E_{N^*}]/E_{NS})$ (see Fig. 1 in SI) for aggregation prone sequences but correlates with $\Delta$ (or equivalently $Z$-score~\cite{lattice}(a)) for foldable sequences. Due to the strong dependence of $\tau_{fib}$  on $P_{N^*}^{max}$ we suggest that only limited number of sequences are aggregation prone (see SI for details).  Although the conclusions were obtained using lattice models they should hold for peptide aggregation.
Our study provides a basis for genome wide search for  consensus sequences with propensity to aggregate.

The work was supported by the Ministry of Science and Informatics in Poland (grant No 202-204-234), grants NSC 96-2911-M 001-003-MY3 \& AS-95-TP-A07, National Center for Theoretical Sciences in Taiwan, and NIH Grant R01GM076688-05.



\begin{widetext}

\begin{center}				  
{\bf \fontfamily{phv} \selectfont Supplementary Information: Determination of factors governing fibrillogenesis of polypeptide chains using lattice models}
\end{center}

For the lattice model with $M$ beads and four types of residues (H, P, +, -), there are $4^M$ sequences. Out of the 65536 (for $M=8$) sequences 5950 have a unique ground state. Among the 65,536 sequences roughly half are mirror images of each other. Although, in reality these would be distinct sequences, in the lattice model they would yield identical thermodynamic properties. Since these sequences are redundant, we do not include them in the statistical analysis. To determine the amylome~\cite{Eisenberg_pnas10_2} (space of sequences in the lattice model that are amyloidogenic) we first determined the folding temperature for the 5950 sequences using the condition $P_{NS}(T_f)=0.5$, where  $P_{NS}(T_f)= e^{-\beta_f E_{NS}}/Z$, where $\beta_f = 1/k_BT_f$, $E_{NS}$ is the energy of the ground state, and $Z = \displaystyle\sum_{i=1}^{p} e^{-\beta E_i}$; $Z$ can be exactly evaluated for each sequence because for $M=8$ the number of conformations, $p$, that satisfy self-avoidance is only 1831. The energies $E_i$ for all the conformations are computed using the chosen interaction energies between the near neighbors beads in the cubic lattice~\cite{MSLi_JCP08_2}. 

Because of the key role that the hairpin-like structure ({\bf N}$^*$) plays in the formation of fibril-like structures  we analyzed, the 5950 sequences with unique ground state (NS) to determine their properties. We determined the temperature $T^*$, where the probability of finding the chain in the {\bf N}$^*$ conformation $P_{N^*}(T)$, is a maximum. We classify those sequences, which satisfy the condition, $1.0 \le T^{*}/T_{f} \le 1.25$ as amyloidogenic. If $T^*$ exceeds $1.25T_f$ the probability of accessing {\bf N}$^*$ is negligible, and such sequences are unlikely to aggregate in finite time scale. Even at $T^{*}=1.25T_f$, many sequences have the maximum in $P_{N^*}(T)$, $P_{N^*}^{max} < 1\%$. In our lattice model there are only 217 such sequences, {\it i.e.}  only 217 (0.66\%) sequences that are aggregation prone. In what follows, we analyze a number of properties of the 217 sequences in order to provide insights into the tendency of natural sequences to be amyloidogenic. 

\bigskip

{\bf Sequence Entropy.}
The sequence entropy for the 217 sequences is calculated using $S_k = -\displaystyle\sum_{i=1}^{4} P_{ki} \ln(P_{ki})$, where $P_{ki}$ is the probability of finding the residue of type $i$ in position $k$. If $P_{ki}=1/4$ (uniform probability) for all $i$ then $S_k = \ln 4 = 1.386$ independent of $k$. We find that $S_k = 1.106, 1.260, 1.329, 1.286, 1.336, 1.318, 1.212$ and 1.064 for $k = 1, 2, ..., 8$ respectively. The terminal positions have a slightly lower entropy than the positions in the middle where the hairpin-like bend is formed. Because there is no position that is strongly conserved we surmise that sequence alone cannot be a good predictor of the tendency of a sequence to aggregate. 
 
\bigskip

{\bf Dependence of $P_{N^*}^{max}$ on $\Delta(=[E_{NS}-E_{N^*}]/E_{NS})$.}
There is striking anit-correlation between $P_{N^*}^{max}$ and $\Delta$ (Fig.~\ref{prob_nstar}), which establishes that in this model the extent of population of the $N^*$ conformation is the major determine of the propensity to aggregate. Interestingly for foldable sequences the $Z$-score (related to $\Delta$ in realistic models) is large. These considerations explain how natural sequences may have evolved to maximize $Z$-score so that unneeded aggregation is avoided. 

\bigskip

{\bf Sequence composition.}
In the 217 sequences, there are 54 different sequence compositions. We find there are different $T^*/T_f$ values for identical composition, which implies that sequence composition alone cannot be a good predictor of the propensity to aggregate\cite{Eisenberg_pnas10_2}.
\bigskip

{\bf Correlation between the type of beads at different positions is significant.}
There is a high degree of correlation between the nature of beads at various positions  (see Tables:~\ref{table1}-\ref{table3}). For example, oppositely charged residues are most likely to be found at positions 1 and 8 (Table:~\ref{table1}). Similarly, if a H bead is in position 2, then with a unit probability, a H bead is found in position 7 (Table:~\ref{table2}). A high preference (0.92) is found for H beads to occupy positions 3 and 6 (Table:~\ref{table3}).  The correlation between the type of beads is not relevant at positions 4 and 5, which are the hairpin bend positions (Table.~\ref{table4}). Based on the results in Tables:~\ref{table1}-\ref{table4} we find that +HHPPHH- is one of the sequences with high probability to aggregate (substantial $P_{N^*}^{max}$).  This sequence has {\bf N}$^*$ as the first excited state, $T^*/T_f=1.169$ and $\Delta =  [E_{NS}-E_{N^*}]/E_{NS} = 0.105$. We also find other sequences with low $T^*/T_f$ values. Three such sequences, which can form ordered fibrils such as the ones shown in Fig.~2a  are ++HPPH- -, +-HPPH+- and +H-PP+H- with identical $\Delta$(=0.10) and $T^*/T_f$(=1.21).

\begin{figure}[h]
\begin{center}
\includegraphics[width=4.0in]{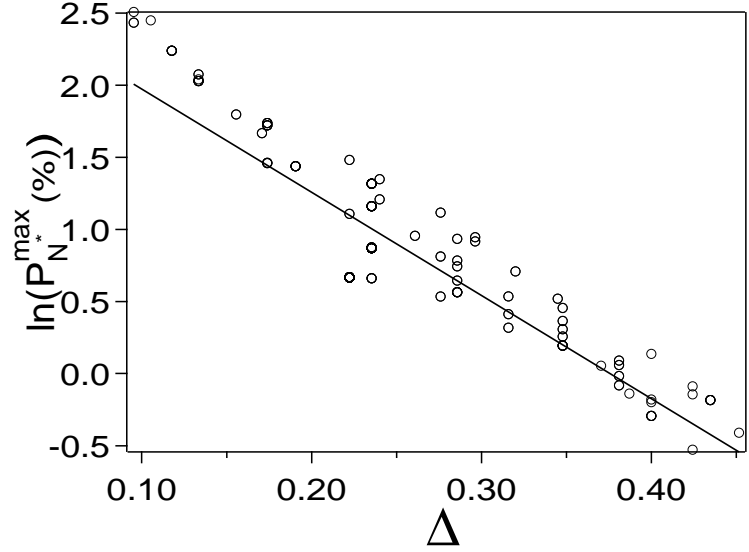}
\caption{Anti-correlation of $\ln(P_{N^*}^{max}$) on $\Delta (= [E_{NS}-E_{N^*}]/E_{NS}$). The straight line $\ln(P_{N^*}^{max}) = -7.15 \Delta + 2.69$ is a fit to the data and has a correlation of 0.9.}\label{prob_nstar}
\end{center}
\end{figure}


\begin{table}[h]
\caption{Bead correlation between positions 1 and 8}
\label{table1}
\begin{center}  
\begin{tabular}{|c|| p{2cm} | p{2cm} | p{2cm} | p{2cm} |}
\hline
\hline
 \multirow{2}{*}{Residue in position-1}  & \multicolumn{4}{|c|}{Probability of finding the residues below in position - 8} \\ \cline{2-5}
   & + & - & H & P \\
\hline
\hline
+ & 0.0 & 0.861  & 0.0 & 0.139 \\ \hline
- & 0.861 & 0.0  & 0.0 & 0.139 \\ \hline
H & 0.0 & 0.0  & 1.0 & 0.0 \\ \hline
P & 0.313 & 0.313 & 0.0 & 0.374 \\ \hline
\hline
\end{tabular}
\end{center}
\end{table}

\begin{table}[h]
\caption{Bead correlation between positions 2 and 7}
\label{table2}
\begin{center}  
\begin{tabular}{|c|| p{2cm} | p{2cm} | p{2cm} | p{2cm} |}
\hline
\hline
 \multirow{2}{*} {Residue in position-2} & \multicolumn{4}{|c|}{Probability of finding the residues below in position - 7 } \\  \cline{2-5}
 & + & - & H & P \\
\hline
\hline
+ & 0.0 & 0.650  & 0.0 & 0.350  \\ \hline
- &  0.650 & 0.0  & 0.0 & 0.350 \\ \hline
H & 0.0 & 0.0  & 1.0 & 0.0 \\ \hline
P & 0.389 & 0.389 & 0.0 & 0.222 \\ \hline
\hline
\end{tabular}
\end{center}
\end{table}

\begin{table}[h]
\caption{Bead correlation between positions 3 and 6}
\label{table3}
\begin{center}  
\begin{tabular}{| c || p{2cm} | p{2cm} | p{2cm} | p{2cm} |}
\hline
\hline
 \multirow{2}{*} {Residue in position-3} & \multicolumn{4}{|c|}{Probability of finding the residues below in position - 6} \\ \cline{2-5}
  & + & - & H & P \\
\hline
\hline
+ & 0.0 & 0.817  & 0.017  & 0.166 \\ \hline
- &  0.817 & 0.0  & 0.017 &  0.166 \\ \hline
H & 0.027 & 0.027  & 0.920 &  0.266 \\ \hline
P & 0.380 & 0.380 & 0.040 & 0.200 \\ \hline
\hline
\end{tabular}
\end{center}
\end{table}

\begin{table}[h]
\caption{Bead correlation between positions 4 and 5}
\label{table4}
\begin{center}  
\begin{tabular}{|c|| p{2cm} | p{2cm} | p{2cm} | p{2cm}|}
\hline
\hline
 \multirow{2}{*}{Residue in position-4} & \multicolumn{4}{|c|}{Probability of finding the residues below in position - 5} \\  \cline{2-5}
  & + & - & H & P \\
\hline
\hline
+ & 0.049 & 0.124  & 0.444 & 0.383 \\ \hline
- & 0.124 & 0.049  & 0.444 &  0.383 \\ \hline
H & 0.336 & 0.336 & 0.085 & 0.243 \\ \hline
P & 0.263 & 0.263 &  0.22 & 0.254 \\ \hline
\hline
\end{tabular}
\end{center}
\end{table}
\end{widetext} 
\end{document}